\documentclass[final,3p,times,twocolumn]{style}

\usepackage{amsmath}
\usepackage{enumitem}
\usepackage{geometry}
\usepackage{amssymb}
\usepackage{tikz}
\usetikzlibrary{arrows.meta, positioning, shapes.geometric}
\usepackage{graphicx}
\usepackage{booktabs}
\usepackage{mathtools}
\usepackage{tabularx}
\usepackage{array}

\begin{document}

\begin{frontmatter}



\title{Scale-Free Priors and Survival Dynamics: A Bayesian Framework for Conflict Duration}

\author{T. F. Stepinski\corref{cor2}}
\ead{stepintz@uc.edu}

\address{Dept. of Geography and GIS, Space Informatics Lab, University of Cincinnati, Cincinnati, OH, USA.}

\cortext[cor2]{Corresponding Author}

\begin{abstract}
We have developed a fully Bayesian survival-analysis framework that reformulates inference about system lifetimes in terms of hazard and survival functions, and extends this representation to interacting actors. Starting from J.~Richard Gott's Copernican principle, we express the scale-free prior as a baseline hazard $\lambda(t)=1/t$, thereby linking a static prior over lifetimes to the dynamic language of survival analysis. In this formulation, Bayesian updating corresponds to conditioning on survival, while the resulting posterior distribution admits a natural representation in terms of hazard and survival functions. The approach is intended for settings where data are sparse or unreliable, and where a scale-free, assumption-light baseline is preferable to heavily parameterized models.

Building on this foundation, we derive general expressions for two-actor systems that characterize joint survival, conditional lifetimes, and comparative outcomes without requiring a specific parametric form of interaction. This yields a flexible and modular framework in which baseline dynamics are separated from interaction effects, allowing different mechanisms to be incorporated transparently. Thus, the primary contribution is a general hazard-based formulation of Bayesian updating and its extension to interacting systems

To illustrate the framework, we consider a multiplicative resource-depletion specification in which interaction modifies the baseline hazard through cumulative engagement intensity. This example demonstrates how interaction terms can be embedded while preserving analytical tractability, including closed-form expressions under simplifying assumptions. We further provide a stylized application to an asymmetric two-actor conflict, the 2026 US/Israel--Iran hostilities, to highlight the qualitative implications of the approach.
\end{abstract}


\end{frontmatter}

\section{Introduction}

Understanding the duration of conflicts has been a central focus of both empirical and theoretical research. Empirical studies typically employ survival-analysis methods, including parametric and semi-parametric hazard models, to estimate how observable factors affect the probability of conflict termination
\citep{fearon1995rationalist, collier2004duration, cunningham2009takes, gates2004modeling, hegre2001toward}. In parallel, a large theoretical literature models conflict as a dynamic bargaining process, in which duration reflects strategic interaction under incomplete information, commitment problems, and learning \citep{fearon1995rationalist, powell2002bargaining, powell2004bargaining}. While these approaches have produced important insights, they typically treat hazard rates either as reduced-form objects to be estimated from data or as outcomes of detailed strategic models requiring strong structural assumptions.

This paper takes a different and complementary approach. Rather than specifying or estimating hazard functions directly, we derive them from a Bayesian formulation of lifetime uncertainty. Starting from Gott’s Copernican principle \citep{gott1993implications}, we adopt a scale-free prior over total lifetime and express it in the language of survival analysis through hazard and survival functions. In this formulation, Bayesian updating corresponds to conditioning on survival, yielding a fully specified baseline hazard without the need for covariates or parametric estimation.

In contrast to the Cox proportional hazards model \citep{cox1972regression}, where covariates act multiplicatively on an otherwise unspecified baseline hazard that must be inferred from data, the present framework derives the entire hazard structure from prior assumptions, with interaction effects introduced explicitly rather than absorbed into regression terms. The framework is particularly suited to settings characterized by limited or unreliable data, absence of a natural time scale, or substantial uncertainty about underlying mechanisms. It therefore provides a transparent, assumption-light benchmark that complements both empirical hazard models and game-theoretic approaches, and serves as a baseline onto which interaction effects can be incorporated.

Gott's argument has been applied to a wide range of phenomena, from the lifetime of Broadway shows to the persistence of political entities \citep{gott1993implications}. However, its original formulation is limited to isolated systems and does not naturally accommodate interactions between multiple actors. In the context of strategic conflict, such interactions---economic pressure, military engagement, resource depletion---are central. Survival analysis has become a standard tool in political science for modeling the duration of conflicts, regimes, and civil wars \citep{emmert2019introduction, whetten2018surviving}, with recent advances in parametric frameworks \citep{crowther2014general} and unified approaches to duration dynamics \citep{chiba2015every} offering powerful methods for empirical analysis. Quantitative modeling of conflict has also highlighted the importance of clear assumptions and the challenges posed by limited data and complex interactions \citep{rexroth2004modeling, yskak2025mathematical}.

The primary objective of this work is to express Bayesian updating in terms of hazard and survival functions and to extend this representation to modeling conflict between two actors. We begin by reformulating Gott's Copernican argument in explicit Bayesian terms, showing how it leads to a scale-free prior and a hazard rate of \(1/t\). We then bridge this static lifetime model to the dynamic perspective of survival analysis, demonstrating that the hazard function is precisely the continuous-time realization of Bayesian updating. 

Building on this foundation, we extend the framework to two interacting actors. First, we derive general expressions for two interacting actors without imposing a specific functional form on the hazard. This yields results that hold for a broad class of models. Then, to illustrate how the general framework can be applied, we consider a specific parametric form in which interaction modifies the hazard multiplicatively through a resource-depletion mechanism. Multiplicative model is especially useful for illustrative purposes because, under the assumption of constant
engagement intensity, it reduces to a semi-analytic formula.

As an illustrative example, we apply the semi-analytic multiplicative variant of the model to the ongoing US/Israel--Iran strategic competition. With realistic parameter choices, the framework produces quantitatively asymmetric predictions consistent with the vast difference in accumulated ``proven survival time'' between the two sides. Throughout, we emphasize the transparency and testability of the approach: all assumptions are explicit, and the model can be readily compared with empirical conflict durations or extended with additional mechanisms.

The remainder of the paper is organized as follows. Section~2 presents the Bayesian formulation of Gott's argument. Section~3 reviews the core concepts of survival analysis and shows how they emerge naturally from the Copernican likelihood. Section~4 extends the framework to two actors, Section~5 introduces describes the resource-depletion multiplicative interaction and its semi-analytic form. Section~6 applies the model to the US/Israel--Iran case. We conclude, in Section 7, with a discussion of the method's strengths and limitations, as well compare the model with Copernican prior to models with exponential and Weibull priors. Table~1 lists symbols used in the paper and their descriptions.

\begin{table}[t]
\centering
\begin{minipage}{1.1\columnwidth}
\caption{Quantities used in the paper}
\label{tab:quantities}

\begin{tabularx}{\linewidth}{lX}
\toprule
\textbf{Symbol} & \textbf{Description} \\
\midrule
\(T\)           & Total lifetime of an actor \\
\(T_{\rm past}\) & Observed age of an actor \\
\(\tau\)        & Additional future survival time \\
\(P(T)\)        & Scale-free prior on total lifetime \\
\(P(T_{\rm past}\mid T)\) & Copernican likelihood \\
\(P(T\mid T_{\rm past})\) & Posterior distribution of lifetime \\
\(f(t)\)        & Prob. density function of lifetime \\
\(S(t)\)        & Survival function \\
\(\lambda(t)\)  & Hazard function \\
\(\Lambda(t)\)  & Cumulative hazard \\
\midrule
\(T_A, T_B\)    & Total lifetimes of actors A and B \\
\(t_A, t_B\)    & Current ages of actors A and B \\
\(P(T_A,T_B\mid t_A,t_B)\) & Joint posterior for two actors \\
\(\lambda_A(t),\lambda_B(t)\) & Hazard rates for actors A and B \\
\(\lambda_{\rm base}(t)=1/t\) & Gott baseline hazard \\
\(\lambda_{\rm conf,A}(t),\lambda_{\rm conf,B}(t)\) & Additive conflict interaction terms \\
\midrule
\(\alpha_A,\alpha_B\) & Vulnerability coefficients \\
\(I(t)\)        & Resource-depletion intensity \\
\(\gamma\)      & Coupling/lethality constant \\
\(E(u)\)        & Instantaneous engagement intensity \\
\(E_0\)         & Constant engagement intensity \\
\(\beta_A=\alpha_A\gamma e_0\) & Effective depletion rate for A \\
\(\beta_B=\alpha_B\gamma e_0\) & Effective depletion rate for B \\
\(\operatorname{Ei}(x)\) & Exponential integral function \\
\(S_A(\tau\mid t_A),S_B(\tau\mid t_B)\) & Conditional survival functions \\
\(P(T_A>T_B\mid t_A,t_B)\) & Probability that A outlasts B \\
\bottomrule
\end{tabularx}
\end{minipage}
\end{table}

\section{Gott's Prior and Its Bayesian Formulation}

In \citep{gott1993implications}, Gott introduced a strikingly simple yet provocative method for predicting the future duration of phenomena ranging from the lifetime of civilizations to the run of Broadway shows. When placed firmly within the Bayesian framework, Gott's argument reveals itself not as a mere heuristics but as a fully specified probabilistic model.

Consider a system---anything from a political regime to a technological civilization---whose total lifetime is an unknown positive quantity \(T\). We observe the system at some past age \(T_{\rm past}\). The hypothesis space consists of all possible values \(T > 0\), and the goal is to infer the posterior distribution of the total lifetime given this single observational datum.

\subsection{The full Bayesian update}

Gott's approach implicitly adopts a \emph{scale-free ignorance prior}:
\begin{equation}
P(T) \propto \frac{1}{T}, \qquad T > 0.
\label{gottPrior}
\end{equation}

This prior expresses complete ignorance about any characteristic timescale: every order of magnitude is treated as equally plausible \emph{a priori}. It is the probabilistic embodiment of the idea that, before seeing any data, there is no reason to favor short lifetimes over long ones or vice versa.

The Copernican principle enters through the likelihood. It asserts that our moment of observation is not special or privileged within the system's total lifetime. Formally, the fraction
\begin{equation}
f = \frac{T_{\rm past}}{T}
\label{gottTime}
\end{equation}
is uniformly distributed on \((0,1)\). Consequently, the likelihood is
\begin{equation}
P(T_{\rm past} \mid T) =
\begin{cases}
\frac{1}{T} & \text{if } 0 < T_{\rm past} < T, \\
0 & \text{otherwise}.
\end{cases}
\label{eq:likelihood}
\end{equation}

Bayes' theorem then combines the prior and likelihood to produce the posterior:
\begin{equation}
P(T \mid T_{\rm past}) = \frac{P(T_{\rm past} \mid T) \, P(T)}{P(T_{\rm past})}.
\label{eq:bayes}
\end{equation}

Substituting the expressions above yields the unnormalized posterior
\begin{equation}
P(T \mid T_{\rm past}) \propto \frac{1}{T^2}, \qquad T \ge T_{\rm past}.
\label{eq:unnormalized-posterior}
\end{equation}

Normalizing gives the elegant closed-form result
\begin{equation}
P(T \mid T_{\rm past}) = \frac{T_{\rm past}}{T^2}, \qquad T \ge T_{\rm past}.
\label{eq:normalized-posterior}
\end{equation}

The components of this Bayesian model can be summarized as follows:

\begin{table}[h!]
\centering
\caption{Bayesian structure of the Copernican (Gott) argument}
\label{tab:copernican_model}
\begin{tabularx}{0.49\textwidth}{|l|X|l|}
\hline
\textbf{Component} & \textbf{Meaning} & \textbf{Expression} \\
\hline
Hypothesis & Total lifetime \(T\) & parameter \\
Prior & Scale-free ignorance (Gott) & \(\propto 1/T\) \\
Evidence & Observed age \(T_{\rm past}\) & data \\
Likelihood & Copernican observation & \(1/T\) \\
Posterior & Updated belief & \(\propto 1/T^2\) \\
\hline
\end{tabularx}
\end{table}

Gott thus supplies both the prior and the key likelihood assumption. Together they produce the posterior that underpins all subsequent predictions. The model is deliberately minimal: it relies on maximal ignorance about timescales and on the symmetry that our observation time is typical. No empirical mechanisms or additional data are required.

\subsection{The 95\% Credible Interval}

The well-known 95\% prediction interval associated with Gott's argument is simply a Bayesian credible interval extracted from the posterior distribution derived above. Starting from
\begin{equation}
P(T \mid T_{\rm past}) = \frac{T_{\rm past}}{T^2}, \qquad T \ge T_{\rm past},
\end{equation}
the cumulative distribution function is
\begin{equation}
F(x) = P(T \le x \mid T_{\rm past}) = 1 - \frac{T_{\rm past}}{x}, \qquad x \ge T_{\rm past}.
\end{equation}
Inverting this CDF gives the quantile function
\begin{equation}
x(p) = \frac{T_{\rm past}}{1-p}.
\end{equation}

For a central 95\% credible interval we take the 2.5\% and 97.5\% quantiles:
\begin{equation}
T_{\rm low} = \frac{T_{\rm past}}{0.975} \approx 1.026\, T_{\rm past},
\end{equation}
\begin{equation}
T_{\rm high} = \frac{T_{\rm past}}{0.025} = 40\, T_{\rm past}.
\end{equation}
Hence the 95\% credible interval is
\begin{equation}
\left[ \frac{T_{\rm past}}{0.975},\; 40\, T_{\rm past} \right].
\end{equation}

The striking asymmetry---a lower bound barely above the observed age and an upper bound forty times larger---arises directly from the heavy-tailed \(1/T^2\) posterior. It reflects that, under maximal ignorance and the Copernican assumption, very long lifetimes remain highly plausible even after a long observed history. This asymmetry is not a flaw; it is a direct consequence of the model's honest admission of uncertainty.

\subsection{Worked Example}

To illustrate the practical use of the model, consider a system observed at age \(T_{\rm past} = 100\) years. The posterior becomes
\begin{equation}
P(T \mid 100) = \frac{100}{T^2}, \qquad T \ge 100.
\end{equation}
The probability that the total lifetime is less than 1000 years is
\begin{equation}
\begin{aligned}
P(T < 1000 \mid 100) &= \int_{100}^{1000} \frac{100}{T^2} \, dT \\
&= 100 \left( \frac{1}{100} - \frac{1}{1000} \right) = 0.9.
\end{aligned}
\end{equation}
Thus there is a 90\% posterior probability that the system will not survive another 900 years. This concrete calculation illustrates how the Bayesian formulation turns Gott's idea into an operational tool.

To summarize, Gott's method is a fully articulated Bayesian model rather than an informal rule of thumb; it corresponds to adopting a scale-invariant prior \(P(T) \propto 1/T\) together with a Copernican likelihood \(P(T_{\rm past} \mid T) = 1/T\), which together produce the posterior \(P(T \mid T_{\rm past}) \propto 1/T^2\). Its special character stems from two features: an extremely weak, scale-invariant prior and a purely symmetry-based likelihood. Because the assumptions are transparent and minimal, the model can serve as a clean baseline---a ``null model''---for more elaborate analyses in which we gradually add real-world mechanisms.

\section{Survival Analysis and Bayesian Updating}

Survival analysis provides a natural framework for modeling the time until an event occurs, such as the failure of a system or the termination of a conflict. Let \(T\) denote the (random) lifetime of a system. Two central objects fully characterize its temporal behavior: the \emph{survival function} and the \emph{hazard function}.

\subsection{Survival and Hazard Functions}

The survival function is defined as
\begin{equation}
S(t) = P(T > t),
\end{equation}
the probability that the system survives beyond time \(t\). It measures how likely the system is to remain ``alive'' (i.e., without experiencing the event) up to time \(t\). Imagine studying how long light bulbs last. Let \(t\) be the time since the bulbs were turned on. At time \(t=100\) hours, some bulbs are still working. Then \(S(100)=0.8\) means that 80\% of the bulbs are still working after 100 hours. A high \(S(t)\) curve indicates that things tend to last a long time; a steep drop means many events happen quickly; a flat region indicates stability (a low-risk period).

The hazard function \(\lambda(t)\) is the instantaneous rate of occurrence of the event at time \(t\), conditional on survival up to that time:
\begin{equation}
\lambda(t) = \lim_{\Delta t \to 0} \frac{P(t \leq T < t+\Delta t \mid T \geq t)}{\Delta t}.
\end{equation}
Intuitively, \(\lambda(t)\) quantifies the immediate risk of failure at time \(t\) given that the system has survived so far.

In the bulbs example, \(\lambda(100)\) tells you the following: among bulbs that have survived up to 100 hours, how likely they are to fail in the next tiny moment. It is not merely a probability---it is a rate: \(\lambda(t)\) equals the probability of failure in a short interval after \(t\), divided by the length of that interval. A large \(\lambda(t)\) means high immediate risk; a small \(\lambda(t)\) means low immediate risk. Increasing \(\lambda(t)\) corresponds to risk that grows with time (e.g., aging machinery or human mortality). Decreasing \(\lambda(t)\) corresponds to risk that is high early and then stabilizes (e.g., infant mortality or early system failures). Constant \(\lambda(t)\) describes a memoryless process (e.g., radioactive decay).

The probability density \(f(t)\), the hazard function \(\lambda(t)\), and the survival function \(S(t)\) are related \citep{govindarajulu2020review} through
\begin{equation}
f(t) = \lambda(t) S(t).
\label{probLambdaSurvivalEq}
\end{equation}

\subsection{Bayesian Updating and Its Representation in Survival Form}

We now describe the relationship between Bayesian updating and the survival-analysis representation used throughout this paper.

\paragraph{Bayesian update as conditioning}
Let $T$ denote the (random) total lifetime of a system, and suppose we observe that the system has survived up to time $T_{\rm past}$. This observation corresponds to the event
\begin{equation}
\{T \ge T_{\rm past}\}.
\label{event}
\end{equation}

Bayesian updating therefore proceeds by conditioning on this event:

\begin{equation}
P(T \mid T_{\rm past})
= P(T \mid T \ge T_{\rm past})
= \frac{P(T)\,\mathbf{1}_{\{T \ge T_{\rm past}\}}}{P(T \ge T_{\rm past})}.
\label{eq:bayes_update_truncation}
\end{equation}

Equivalently, in proportional form,
\begin{equation}
P(T \mid T_{\rm past}) \propto P(T)\,\mathbf{1}_{\{T \ge T_{\rm past}\}}.
\label{eq:bayes_update_truncation_proportional}
\end{equation}
In this formulation, the likelihood $P(T_{\rm past}\mid T)$ takes the form of an indicator function:

\begin{equation}
P(T_{\rm past}\mid T) = \mathbf{1}_{\{T \ge T_{\rm past}\}},
\label{indicator}
\end{equation}
which reflects the fact that only lifetimes compatible with the observation are retained. Thus, the Bayesian update is a truncation (or selection) of the prior distribution followed by renormalization.

\paragraph{Rewriting the prior in survival form}
Substituting \eqref{probLambdaSurvivalEq} into \eqref{eq:bayes_update_truncation} yields
\begin{equation}
P(T \mid T_{\rm past})
= \frac{\mathbf{1}_{\{T \ge T_{\rm past}\}}\, \lambda(T)\, S(T)}
{\displaystyle \int_{T_{\rm past}}^\infty \lambda(u)\, S(u)\,du}.
\label{eq:posterior_hazard_form}
\end{equation}

\paragraph{Conditional survival and posterior factorization}
For $T \ge T_{\rm past}$, the conditional survival function is
\begin{equation}
S(T \mid T_{\rm past}) = \frac{S(T)}{S(T_{\rm past})}.
\label{eq:conditional_survival}
\end{equation}
Using this identity, the posterior density \eqref{eq:posterior_hazard_form} can be written in the compact form
\begin{equation}
P(T \mid T_{\rm past}) = \lambda(T)\, S(T \mid T_{\rm past}), 
\qquad T \ge T_{\rm past}.
\label{eq:posterior_lambda_S}
\end{equation}

\paragraph{Interpretation}
Equation \eqref{eq:posterior_lambda_S} expresses the posterior density as the product of two terms:
\begin{itemize}
\item $S(T \mid T_{\rm past})$: the probability that the system survives up to time $T$, given that it has already survived to $T_{\rm past}$;
\item $\lambda(T)$: the instantaneous rate at which termination occurs at time $T$.
\end{itemize}
Thus, the posterior density at time $T$ is obtained by combining the probability of surviving up to $T$ with the instantaneous risk of termination at that moment.

\paragraph{Clarification.}
It is important to distinguish between the \emph{updating step} and its \emph{representation}. The Bayesian update occurs through conditioning on the event $\{T \ge T_{\rm past}\}$, as in \eqref{eq:bayes_update_truncation}. Equation \eqref{eq:posterior_lambda_S} does not constitute a new updating rule; rather, it is a re-expression of the resulting posterior density in terms of hazard and survival functions. This representation is useful because it links Bayesian inference to the standard structure of survival analysis and facilitates interpretation in terms of risk and persistence over time.

\subsection{Example: Scale-Free (Gott) Model}

This example demonstrates how the posterior formula in Gott's model (Eq.~(\ref{eq:normalized-posterior}) in Section~2) can be derived from Bayesian updating expressed in terms of hazard and survival functions (Eq.~(\ref{eq:posterior_lambda_S})).

In Gott’s model, the risk of ending a process at a given moment is inversely related to how long the system has already lasted (scale-free lifetime), and the observation is assumed to occur at a typical, non-special time point (Copernican principle). The larger the hypothetical total lifetime \(T\), the less likely it is that your randomly timed observation would have occurred as early as \(T_{\rm past}\). Thus, the survival function is \(S(T)=T_{\rm past}/T\) and the hazard function is \(\lambda(T)=1/T\). Substituting these Gott-model-specific functions of hazard and survival into Eq.~(\ref{eq:posterior_lambda_S}) yields \(P(T \mid T_{\rm past}) = T_{\rm past}/T^2\) for \(T \ge T_{\rm past}\), recovering Eq.~(\ref{eq:normalized-posterior}).

\section{Extension to Two-Actor Conflict}

The modular structure developed in Section~3 naturally extends to the modeling of conflict between two actors (e.g., countries, political regimes, or competing systems). Let \(T_A\) and \(T_B\) denote the total lifetimes of actors A and B, respectively. We observe both actors still active at elapsed times \(T_{\rm past}^A\) and \(T_{\rm past}^B\) (the current ages of the two actors). The goal is to obtain the joint posterior distribution \(P(T_A, T_B \mid T_{\rm past}^A, T_{\rm past}^B)\) and derived quantities such as the probability that one actor outlasts the other.

The construction of the two-actor model follows exactly the same logic as used earlier. The scale-free ignorance priors remain unchanged:
\begin{equation}
P(T_A) \propto \frac{1}{T_A}, \quad T_A > 0; \qquad
P(T_B) \propto \frac{1}{T_B}, \quad T_B > 0.
\label{eq:priors-two}
\end{equation}
The Copernican baseline hazard for each actor is still \(\lambda_{\rm base}(t) = 1/t\).

Interaction between the two actors enters solely through two conflict terms added to the baseline hazard:
\begin{equation}
\lambda_A(t) = \frac{1}{t} + \lambda_{\rm conf,A}(t;\theta), \qquad t \ge T_{\rm past}^A,
\label{eq:hazard-A}
\end{equation}
\begin{equation}
\lambda_B(t) = \frac{1}{t} + \lambda_{\rm conf,B}(t;\theta), \qquad t \ge T_{\rm past}^B.
\label{eq:hazard-B}
\end{equation}
Here \(\theta\) collects any parameters describing the conflict (interaction strength, asymmetry, escalation rate, dependence on the other actor’s age, etc.). The functions \(\lambda_{\rm conf,A}\) and \(\lambda_{\rm conf,B}\) may be chosen freely; the model remains fully modular.

The joint posterior density follows directly from Eq.~(\ref{eq:posterior_lambda_S}):
\begin{equation}
\begin{split}
P(T_A, T_B \mid T_{\rm past}^A, T_{\rm past}^B)
&= \lambda_A(T_A)\,\lambda_B(T_B)\, \\
&\quad S(T_A, T_B \mid T_{\rm past}^A, T_{\rm past}^B)
\end{split}
\label{eq:joint-posterior}
\end{equation}
for \(T_A \ge T_{\rm past}^A\), \(T_B \ge T_{\rm past}^B\) (and zero otherwise).

From Eq.~(\ref{eq:joint-posterior}) one can compute the quantities of primary interest in a conflict setting:
\begin{itemize}
\item Marginal posteriors for each actor’s total lifetime by integrating out the other variable.
\item The probability that actor A outlasts actor B:
\begin{equation}
\begin{split}
P(T_A > T_B \mid T_{\rm past}^A, T_{\rm past}^B) \\
= \iint_{T_A > T_B} P(T_A, T_B \mid T_{\rm past}^A, T_{\rm past}^B)\, dT_A\, dT_B.
\end{split}
\label{eq:prob-A-outlasts-B}
\end{equation}
\item Expected remaining lifetimes \(\mathbb{E}[T_A - T_{\rm past}^A \mid T_{\rm past}^A, T_{\rm past}^B]\) and \(\mathbb{E}[T_B - T_{\rm past}^B \mid T_{\rm past}^A, T_{\rm past}^B]\).
\item The posterior distribution of the lifetime difference \(T_A - T_B\).
\end{itemize}

All of these quantities are obtained by standard integration (analytic for simple choices of \(\lambda_{\rm conf,A}\) and \(\lambda_{\rm conf,B}\), numerical otherwise). The model requires no additional machinery beyond the survival-analysis and continuous-time Bayesian updating framework: the Gott baseline is encoded once and for all in the \(1/t\) terms, while all conflict dynamics enter exclusively through the two interaction hazards. This preserves the transparency and modularity emphasized throughout the paper.

Useful visualizations include marginal survival curves for each actor, contour plots of the joint posterior, the probability \(P(T_A > T_B)\) as a function of conflict parameters, and the posterior density of the lifetime difference. These plots reveal how interaction strength and asymmetry deform the baseline Copernican predictions and quantify the advantage (or disadvantage) conferred by the conflict dynamics.

The formulation thus provides a clean, fully Bayesian, survival-analysis-based template for modeling strategic conflict between two actors while remaining faithful to the minimal assumptions of Gott’s original argument.

\section{Resource-Depletion Multiplicative Interaction}

In equations (\ref{eq:hazard-A}) and (\ref{eq:hazard-B}) the combination of hazards due to the prior (baseline hazard) and due to interaction between the two actors is additive. An additional hazard due to interaction contributes to an absolute increase in risk, independent of the current baseline level. Another possibility is to use the resource-depletion multiplicative interaction, where the hazard is modeled as a multiplicative process \citep{cox1972regression}.

\begin{equation}
\lambda_A(t) = \frac{1}{t} \cdot \exp\bigl(\alpha_A \, I(t)\bigr), \qquad t \ge t_A,
\label{eq:hazard-A-multi}
\end{equation}
and likewise for actor B:
\begin{equation}
\lambda_B(t) = \frac{1}{t} \cdot \exp\bigl(\alpha_B \, I(t)\bigr), \qquad t \ge t_B.
\label{eq:hazard-B-multi}
\end{equation}

Here \(\alpha_A > 0\) and \(\alpha_B > 0\) are actor-specific vulnerability coefficients (larger values mean greater sensitivity to depletion), and the shared resource-depletion intensity \(I(t)\) is defined as
\begin{equation}
I(t) = \gamma \int_{t_0}^{t} E(u) \, du,
\label{eq:I-depletion}
\end{equation}
where the integration is over calendar time from \(t_0\) (the start of the observable conflict) to the future time \(t\). \(E(u)\) is the instantaneous engagement intensity (rate of battles, economic pressure, or resource expenditure at time \(u\)), and \(\gamma > 0\) is a coupling/lethality constant common to both actors. In the resource-depletion multiplicative model the interaction between actors enters as a ``scaling'' factor \(\exp(\alpha I(t))\) that multiplies the baseline hazard. The integral runs from \(t_0\) to \(t\) because depletion is a shared, time-dependent environmental process that accumulates over the entire duration of the conflict, not over each actor’s individual remaining lifetime.

The joint posterior density is still given by Eq.~(\ref{eq:joint-posterior}); only the functional form of \(\lambda_A\) and \(\lambda_B\) is updated by substituting Eqs.~(\ref{eq:hazard-A-multi})--(\ref{eq:hazard-B-multi}). All derived quantities—marginal posteriors, \(P(T_A > T_B \mid t_A, t_B)\) (Eq.~(\ref{eq:prob-A-outlasts-B})), expected remaining lifetimes, and the distribution of \(T_A - T_B\)—follow from the same integration as before.

Mathematically, in the additive model the hazard appears as the linear sum of baseline and interaction hazards, while in the multiplicative model the interaction hazard appears inside the exponent of the survival function (via the cumulative hazard \(\int \lambda(u)\,du\)). Consequently, the additive model produces a linear shift in the cumulative hazard, whereas the multiplicative model produces an accelerating risk that grows with the integral of engagement; the older a system already is (larger \(t\)), the more strongly the depletion amplifies its instantaneous termination probability. This makes the multiplicative resource-depletion model easier to interpret for sustained attrition conflicts (the longer the fight lasts, the faster both sides exhaust), while the additive model is simpler for sudden, memoryless shocks. Both remain fully compatible with the Copernican baseline and the continuous-time Bayesian updating machinery developed earlier; the choice between them is a modeling decision that can be tested by comparing posterior predictions against observed conflict duration.

\subsection{Semi-Analytic Survival Formulas for Constant Engagement Intensity}

A particularly tractable case of the resource-depletion multiplicative model occurs when the engagement intensity is constant, \(E(t)=E_0\). In this situation the depletion intensity becomes linear in time:
\begin{equation}
I(t)=\gamma E_0(t-t_0).
\label{eq:I-constant}
\end{equation}

Let us define the effective depletion rates
\begin{equation}
\beta_A=\alpha_A\gamma E_0,\qquad\beta_B=\alpha_B\gamma E_0.
\label{eq:beta-def}
\end{equation}

The hazard functions then simplify to
\begin{equation}
\lambda_A(t)=\frac{1}{t}\exp\bigl(\beta_A(t-t_0)\bigr),\qquad t\ge t_A,
\label{eq:lambda-A-constant}
\end{equation}
and likewise for actor B.

The cumulative hazard \(\Lambda_A\) for actor A from the current age \(t_A\) to a future time \(t_A + \tau\), where \(\tau\) is additional future survival time, is therefore
\begin{equation}
\Lambda_A(\tau)=\int_{t_A}^{t_A+\tau}\frac{\exp\bigl(\beta_A(u-t_0)\bigr)}{u}\,du.
\label{eq:cumhaz-A}
\end{equation}

This integral can be expressed exactly in terms of the exponential integral function \(\operatorname{Ei}(x)\) \citep{debelu2024new}:
\begin{equation}
\Lambda_A(\tau)=\operatorname{Ei}\bigl(\beta_A (t_A + \tau - t_0)\bigr)-\operatorname{Ei}\bigl(\beta_A(t_A-t_0)\bigr).
\label{eq:Lambda-Ei}
\end{equation}

Consequently, the conditional survival function for actor A is
\begin{equation}
\begin{split}
S_A(\tau\mid t_A) &= \exp\bigl(-\Lambda_A(\tau)\bigr) \\
&= \exp\Bigl(-\bigl[\operatorname{Ei}(\beta_A(t_A+\tau-t_0))+\operatorname{Ei}(\beta_A(t_A-t_0))\bigr]\Bigr).
\end{split}
\label{eq:SA-analytic}
\end{equation}
An analogous expression holds for \(S_B(\tau\mid t_B)\).

Because the depletion intensity \(I(t)\) is a function of calendar time only, the two lifetimes are conditionally independent given the observed ages. The joint conditional survival function is therefore simply the product
\begin{equation}
S(\tau_A,\tau_B\mid t_A,t_B)=S_A(\tau_A\mid t_A)\,S_B(\tau_B\mid t_B).
\label{eq:joint-S-analytic}
\end{equation}

The joint posterior density follows immediately from the fundamental identity already established in the paper (Eq.~(\ref{eq:joint-posterior})):
\begin{equation}
\begin{split}
P(T_A,T_B\mid t_A,t_B) \\
= \lambda_A(T_A)\,\lambda_B(T_B)\,S_A(T_A-t_A\mid t_A)\,S_B(T_B-t_B\mid t_B).
\end{split}
\label{eq:joint-posterior-analytic}
\end{equation}

\section{Example: US/Israel vs Iran 2026 conflict}

We analyze a specific two-actors conflict using the methodology described in sections 2 to 5. Specifically, we use the resource-depletion multiplicative interaction model and assume the constant engagement intensity \(E(t)=E_0\). Thus, the results are obtained using semi-analytic survival formulas given in section 4.2. 

The choice of parameters for this example is inspired by the US/Israel--Iran conflict which erupted on February, 28 2026 with large-scale airstrikes by the United States and Israel targeting Iranian leadership, nuclear and ballistic-missile facilities, military command centers, and air defenses. Iran responded with missile and drone attacks on Israeli and US positions, while Hezbollah intensified operations from Lebanon. Iran also closed the Strait of Hormuz, severely disrupting global oil flows. 

\subsection{Criteria for Selecting Parameter Values}

Since, as of the writing of this paper in April 2026, there is no direct empirical data on the ongoing (or hypothetical future) US/Israel–Iran conflict that would allow statistical fitting of the parameters, the values of $  \alpha_A  $, $  \alpha_B  $, $  \gamma  $, and $  E_0  $ are chosen using qualitative, structural, and consistency-based criteria rather than data-driven estimation. The goal is to produce realistic, interpretable, and competitive short-term behavior (first 100 weeks) while preserving the long-term age advantage of actor A (US/Israel)

\begin{figure*}[t]
	\includegraphics[width=17cm]{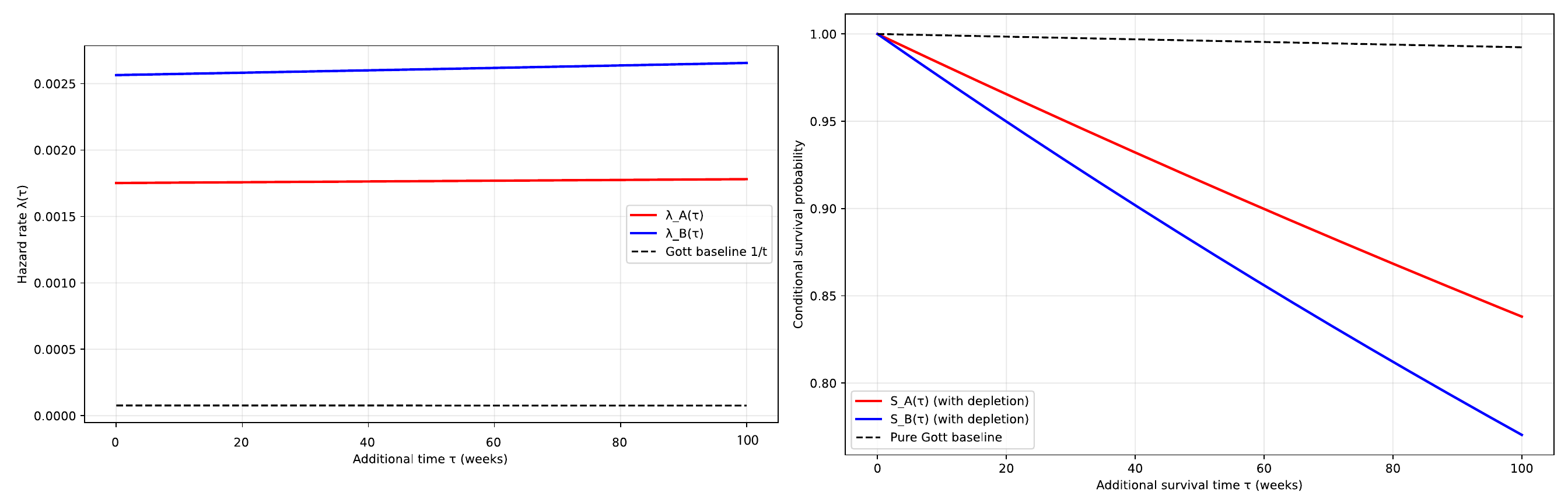}
\caption{Resource-depletion multiplicative interaction model of conflict between actors A (US/Israel) and B (Iran) with parameters  $\gamma = 0.015 $,  $E_0 = 2.0 $,  $ \alpha_A = 0.009 $, and $ \alpha_B = 0.026$. (Left) Hazard trajectories. (Right) Survival curves.}
\label{fig1}
\end{figure*}

 The following qualitative and structural criteria are used to chose the parameters:

\begin{itemize}
\item {\sl Qualitative asymmetry in resilience.}  
  US/Israel (actor A) has vastly greater accumulated resources, technological superiority, alliances, economic depth, and institutional resilience compared to Iran (actor B).  
  Therefore: \(\alpha_A \ll \alpha_B\) (estimated \(\alpha_A\) is \(1/3\) to \(1/4\) of \(\alpha_B\)).

\item {\sl Competitiveness on short timescales (first 50--100 weeks)}  
  The model should show a noticeable but not catastrophic difference in survival probabilities.  
  \(\alpha_B\) is set high enough that depletion has a visible effect on Iran, while \(\alpha_A\) is kept low enough that US/Israel remains relatively resilient.

\item {\sl Moderate overall depletion rate}  
  \(\gamma\) (coupling/lethality) and \(E_0\) (engagement intensity) are chosen so that the cumulative depletion \(I(t) = \gamma e_0 t\) grows at a plausible rate.  
  Too low \(\to\) almost no effect; too high \(\to\) unrealistic collapse of both sides within weeks.

\item {\sl Consistency with the model structure}  
  The parameters must preserve the Gott baseline (\(1/t\)) as the dominant long-term effect.  
  The depletion term should modulate but not override the age difference (\(t_A \gg t_B\)) on short timescales.
\end{itemize}

Using these criteria we set parameters to the following values: $ t_A = 13{,}035 $ weeks (about 250 years, the age of the US), $ t_B = 2{,}451 $ weeks (about 47 year, the age of the Islamic Republic of Iran), $\gamma = 0.015 $ – 0.035,  $E_0 = 1.0 $ – 2.0, $ \alpha_A = 0.008 $ – 0.011, and $ \alpha_B = 0.025$ – 0.032 (Iran). These values produce survival probabilities after 100 weeks in the range 0.85–0.92 for A and 0.78–0.85 for B, which is plausible for a sustained but not all-out global conflict. The values of parameters are illustrative rather than calibrated. Their purpose is to explore the qualitative consequences of the model (strong age advantage vs. depletion-driven vulnerability) rather than to make precise predictions. They can be adjusted for sensitivity analysis or to explore different conflict intensities.

Fig.1 shows how values of hazard and survival functions for both actors change with survival after the start of the conflict time $\tau$. The left panel  shows the hazard rates $  \lambda_A(\tau)  $ and $  \lambda_B(\tau)$  together with the pure Gott baseline hazard for actor A as functions of additional time $  \tau  $ from the start of the conflict to $\tau =100$ weeks. The Y-axis is the instantaneous termination rate per week.

The Gott baseline $  \lambda(t) = 1/t  $ is inherently low when the current age $  t  $ is large. At $  t \approx 13{,}000  $ weeks the baseline is on the order of $  10^{-4}  $ per week or smaller. This reflects the core Copernican idea: the longer a system has already survived, the lower its instantaneous risk of termination at any given moment. The depletion terms, $\exp\bigl(\alpha \, I(t)\bigr)$, add only a modest increase on top of this already tiny baseline, so the hazard rates show very little increase on 100 weeks timescale which is reflected in near-flatness of both hazard trajectories. The separation between the two hazard curves demonstrates the asymmetric impact of depletion. This behavior aligns well with the qualitative expectation that the older, more resilient actor (US/Israel) remains relatively stable in the first 100 weeks, while the younger actor (Iran) faces gradually increasing risk.

The right panel on Fig.1 shows the conditional survival probabilities $  S_A(\tau)  $ and $  S_B(\tau)  $ over the first 100 weeks of  conflict time $\tau$, together with the pure Gott baseline for actor A. The Y-axis is the conditional survival probability. Both survival curves decrease noticeably over 100 weeks, but actor A  decreases more slowly than actor B due to actor B having a higher hazard rate (blue line in the hazard plot), which directly translates into a faster drop in survival probability. At $  \tau = 100$ weeks  $S_A \approx 0.84 $ (US/Israel retains fairly high survival probability) and $ S_B \approx 0.77 $ (Iran’s survival probability drops more significantly). The pure Gott baseline (black dashed line) is almost flat and remains very high ($\sim$0.99), confirming that without any depletion both actors would have extremely high survival probability on this timescale.

Overall, Fig.1 demonstrates a moderate but clear asymmetry in the short term. US/Israel (actor A) is noticeably more resilient to depletion, maintaining higher survival probability. Iran (actor B) is more sensitive to the cumulative effect of engagement, leading to a faster erosion of its survival outlook. The depletion is strong enough to create a visible difference, but not so extreme that either side collapses after the first 100 weeks.

\begin{table*}[t]
\renewcommand{\arraystretch}{1.3}
\caption{Comparison of Survival Predictions: Copernican vs. Traditional Models}
\label{table:model_comparison}
\centering
\begin{tabular}{|p{3.5cm}|p{4cm}|p{4cm}|p{4cm}|}
\hline
\textbf{Feature} & \textbf{Copernican Model} & \textbf{Exponential Model} & \textbf{Weibull Model} \\ \hline
\textbf{Hazard Assumption} & \textbf{Decreasing ($\lambda(t) \propto 1/t$):} The longer a regime exists, the lower its daily risk. & \textbf{Constant ($\lambda$):} Risk is the same on Day 1 as it is on Day 1,000 (Memoryless). & \textbf{Flexible:} Risk can steadily increase (aging/wear) or decrease over time. \\ \hline
\textbf{Iran Prediction (Age: $\approx$ 47 yrs)} & \textbf{High Vulnerability:} As the younger actor, its ``survival capital'' erodes faster under depletion. & \textbf{Static Risk:} Survival depends solely on current intensity, ignoring its 47-year history. & \textbf{Moderate:} Often predicts increasing risk over time if the shape parameter $\rho > 1$. \\ \hline
\textbf{US Prediction (Age: $\approx$ 250 yrs)} & \textbf{High Resilience:} Its vast ``proven survival time'' acts as a massive buffer against termination. & \textbf{Identical Baseline:} Without different $\lambda$ values, it is treated as equally risky as Iran. & \textbf{Variable:} Can model the US as more stable, but requires ``fitting'' to historical data. \\ \hline
\textbf{95\% Credible Interval} & \textbf{Extremely Wide:} Predicts a 95\% chance of ending between $\approx 1$ and 40 times its past age. & \textbf{Tighter:} Based on the mean survival time ($\mu = 1/\lambda$). & \textbf{Narrowest:} Most precise if the underlying data distribution is actually known. \\ \hline
\end{tabular}
\end{table*}

\section{Discussion and Conclusion}

This paper has developed a fully Bayesian survival-analysis framework for modeling strategic conflict between two actors that is directly grounded in J.~Richard Gott's Copernican principle. By reformulating Gott's scale-free prior \(P(T) \propto 1/T\) as a hazard function \(\lambda(t) = 1/t\), we have shown that the Copernican argument is precisely the continuous-time realization of Bayesian updating. The hazard and survival functions thereby provide a natural bridge between Gott's static lifetime model and the dynamic language of survival analysis.

\subsection{Discussion of the proposed framework}

The framework is deliberately modular. The Copernican baseline \(\lambda(t) = 1/t\) is preserved as the dominant long-term effect for any actor, while all interactions between the two actors enter exclusively through conflict terms. We emphasize that the results derived in Section 4 do not depend on a particular parametric specification of the hazard; the multiplicative interaction model considered in Section 5 serves only to illustrate how interaction effects may be incorporated in practice.

Applying these semi-analytic expressions to the ongoing US/Israel--Iran conflict (as of April 2026), the model produces quantitatively asymmetric yet realistic short-term predictions. With parameters chosen on the basis of qualitative resilience differences (\(\alpha_A \ll \alpha_B\)) and moderate depletion rates, Actor~A (US/Israel) retains substantially higher survival probability over the first 100 weeks than Actor~B (Iran), while both curves remain competitive. This asymmetry arises naturally from the large current-age difference (\(t_A \gg t_B\)) and the cumulative nature of resource depletion, without any additional mechanistic assumptions. It is important to note that, as the conflict progresses, new information about each actor's resilience (or lack thereof) may become available, which could lead to a reassessment of the chosen parameter values.

The approach differs fundamentally from conventional parametric survival models used in conflict research. Exponential (and Weibull priors introduce explicit characteristic timescales and are typically chosen for mechanistic or data-fitting reasons \citep{crowther2014general, collier2004duration, fearon2004some, box2004event, chiba2015every}. In contrast, Gott's scale-free prior imposes no preferred timescale and serves as a transparent null model onto which interaction terms can be added in a modular fashion.

\subsection{Discussion of the example}

Because the conflict is still "young" (under two months as of the writing of this paper), the proposed Copernican model is highly sensitive. It yields a 95\% credible interval suggesting the conflict could end as early as tomorrow or last for up to 5.7 years. Thus, it refuses to rule out a "forever war" simply because there is not enough history to prove otherwise. On the other hand, if we assume a "typical" modern air campaign duration (e.g., a mean of four weeks), an exponential model would suggest a constant $\approx$25\% chance of the conflict ending each week. Such a model fails to account for the fact that the April 8, 2026, ceasefire has already shifted the conflict's "age" into a more stable, albeit fragile, phase. A Weibull model with $\beta <1$ (decreasing hazard) would suggest that since the conflict survived the initial "Epic Fury" strikes, it is now more likely to settle into a prolonged, lower-intensity stalemate.

Table 3 compares features of the Copernican, exponential (Compertz model \citep{kirkwood2015deciphering}), and Weibull models in the context of the US/Israel–Iran conflict. Note that the term "risk" is broadly defined; in this context, it generally refers to the conclusion of the conflict in a manner where a given actor is perceived as the losing side.

\subsection{Conclusion}

The principal strengths of the framework are its analytical transparency, minimal assumptions, and interpretability. All parameters have clear substantive meaning (vulnerability coefficients, coupling/lethality, engagement intensity), and the model can be extended with time-varying engagement, more actors, or empirical calibration as data become available. Limitations include the illustrative nature of the current parameter choices and the assumption of constant engagement intensity in the semi-analytic case. Future work can address these by incorporating stochastic or time-dependent \(E(t)\) and by confronting the model with historical conflict-duration data.

By uniting Gott's Copernican insight with survival analysis in a fully Bayesian setting, this paper provides a new, minimalist yet extensible template for modeling strategic conflict. It demonstrates that powerful, mechanism-light predictions can be obtained even in the absence of detailed internal data, opening a complementary path alongside traditional mechanistic and econometric approaches.


\end{document}